\documentclass[11pt]{article}
\usepackage{moriond,epsfig}

\begin{document}

\vspace*{4cm}
\title{REVIEW OF HIGGS PHYSICS AT THE LHC}

\author{ VOLKER DROLLINGER }

\address{ Dipartimento di Fisica "Galileo Galilei", Universit\a`a di Padova,\\
Via Marzolo 8, 35131 Padova, Italy}

\maketitle\abstracts{
Nobody knows exactly what kind of Higgs physics will be unveiled when the Large Hadron Collider is turned on. There could be one Standard Model Higgs boson or five Higgs bosons as is the case in two-Higgs-doublet models; there could be more exotic or even completely unexpected scenarios. In order to be prepared for the LHC era, a solid understanding of Standard Model or Standard-Model-like Higgs physics is necessary. The first goal is to discover the Higgs boson. Afterwards it has to be proven that the new particle is indeed a Higgs boson. The Higgs boson has to couple to mass and its spin has to be zero. Additional observables, such as decay width or CP eigenvalue, help to distinguish between different models. Due to an almost infinite variety of models, another important goal is to prepare for all possible situations. For example, Higgs bosons could be produced in decays of heavier particles, or could decay to invisible particles. In the following, a selection of mainly new studies by ATLAS and CMS is presented.}

\section{Introduction}
A new era of Higgs boson hunting will start at the LHC (Large Hadron Collider) which will collide protons with protons. With a centre-of-mass energy of 14 ${\rm TeV}$ and a design luminosity of ${\rm L = 10^{34}\ cm^{-2}s^{-1}}$, the LHC will be able to produce plenty of Standard Model Higgs bosons with a rate sufficient for detection up to the triviality limit of about 1 ${\rm TeV/}c^2$.

\begin{figure}
  \centering
  \vspace*{-4mm}
  \epsfig{figure=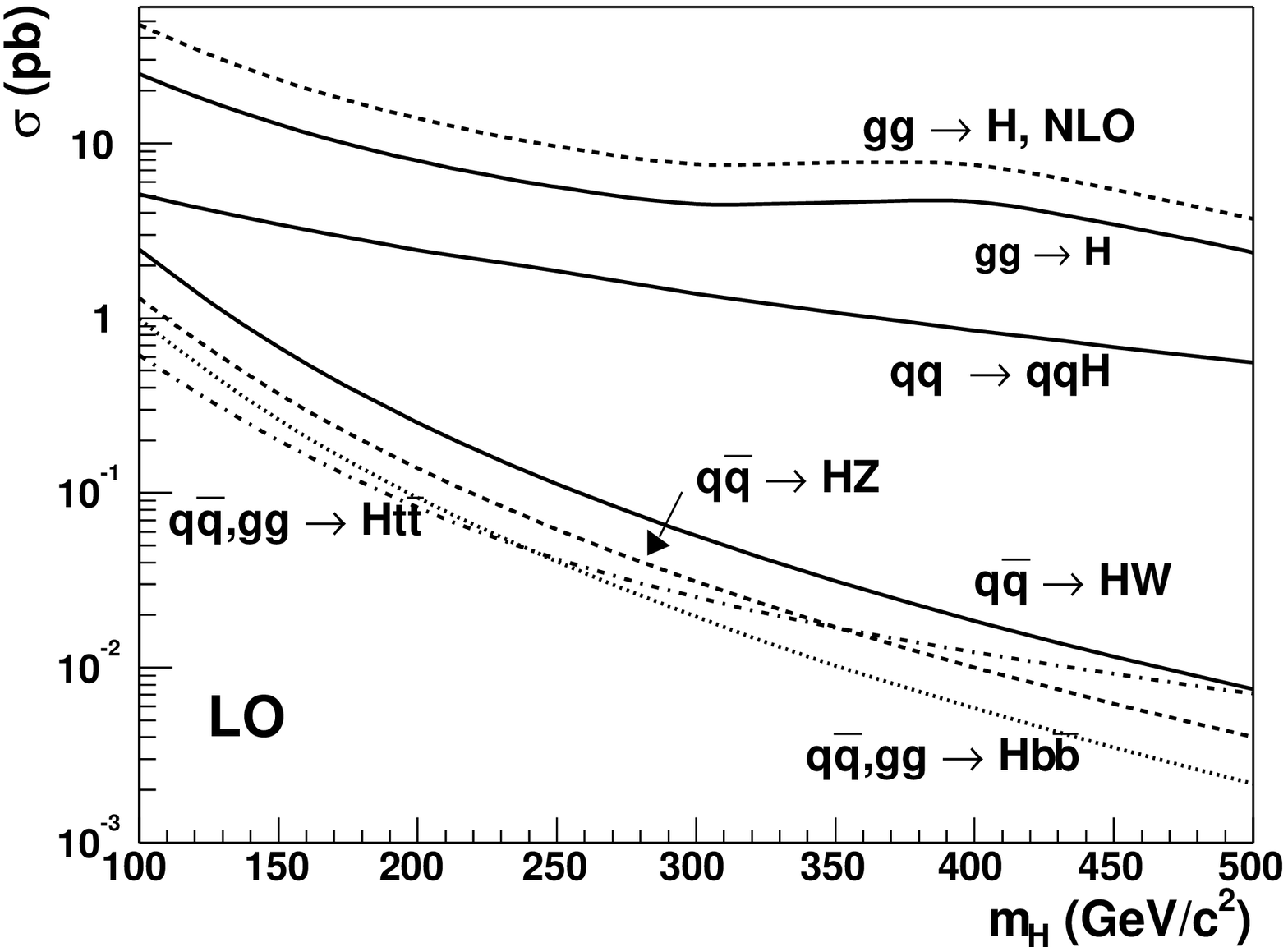,width=0.45\textwidth}
  \epsfig{figure=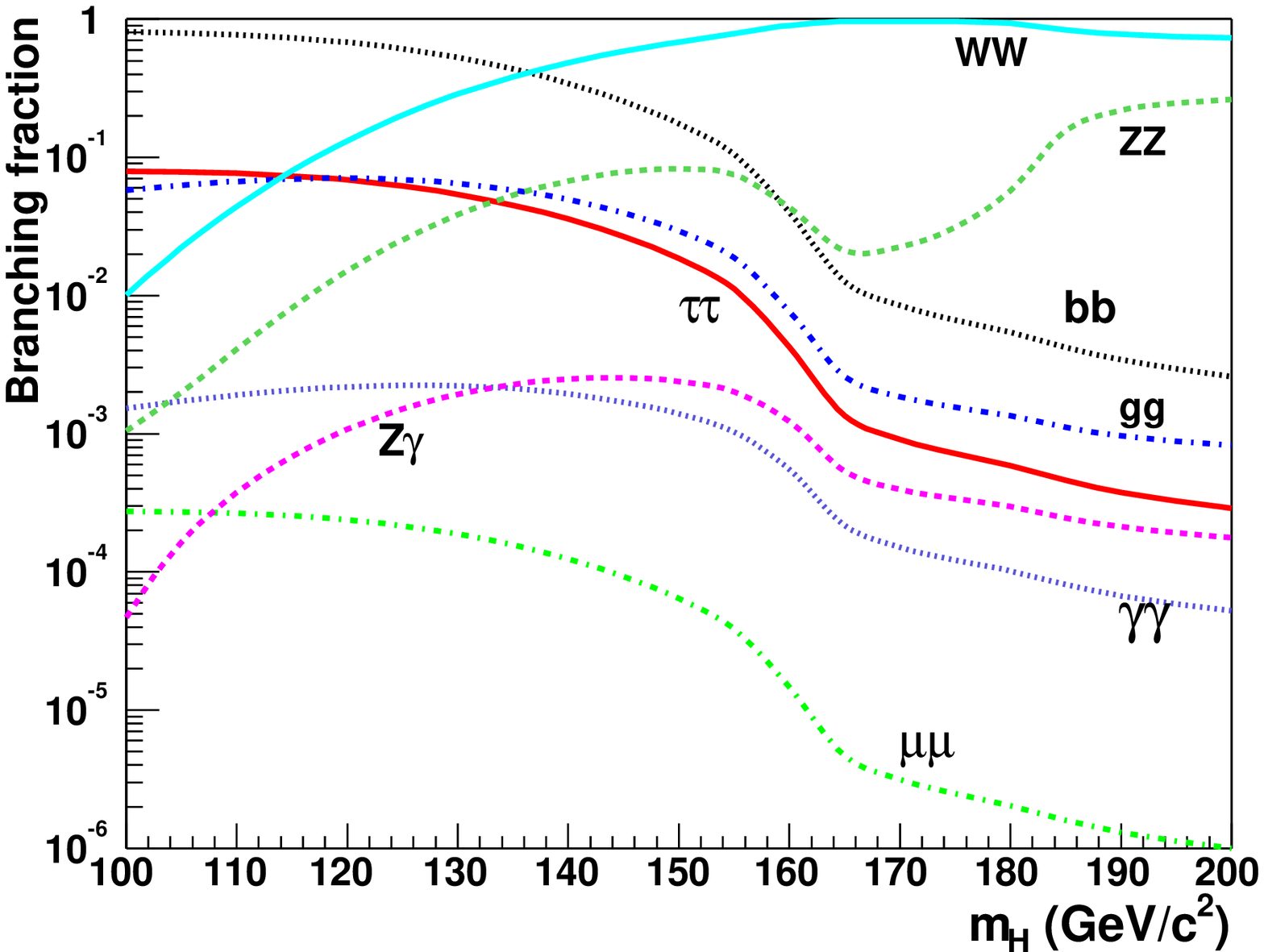,width=0.45\textwidth}
  \put(-385,123){\scalebox{1.1}[1.1]{a)}}
  \put(-180,123){\scalebox{1.1}[1.1]{b)}}
  \vspace*{-0.5mm}
  \caption{a) Cross sections for Standard Model Higgs boson production at the LHC. b) Branching ratios for decays of the Standard Model Higgs boson.
  \label{fig:sm_higgs}}
  \vspace*{-0.5mm}
\end{figure}

Even in the Standard Model with only one Higgs boson, there are several possibilities to produce this particle at the LHC~\cite{ref:h_prod}$^,\ $\cite{ref:hdecay}, as illustrated in Fig.~\ref{fig:sm_higgs}a. The most important production processes are gluon fusion, vector boson fusion, and associated production with top quarks or vector bosons. Depending on the Higgs boson mass, different decay modes open up, as shown in  Fig.~\ref{fig:sm_higgs}b. At low mass, the decay to ${\rm b\bar{b}}$ is dominant, but the decay ${\rm h \rightarrow \tau^+ \tau^-}$ is also sizeable. With increasing mass, the Higgs boson preferably decays to heavier particles, mainly to ${\rm Z Z}$ and ${\rm W^+ W^-}$ pairs. The branching ratio of the ${\rm h \rightarrow \gamma \gamma}$ decay is only at the level of one permil or below, but can be exploited below the ${\rm W^+ W^-}$ threshold nevertheless because of its distinct final state. The production processes in combination with the possible decay channels result in a large number of possible final states. In addition, decays of top quarks, W bosons and Z bosons increase this number. Detectors that can measure accurately the particles of all the final states, such as photons or leptons, are needed. In the following, electrons and muons are referred to as leptons.

The ATLAS (A Toroidal LHC ApparatuS) and CMS (Compact Muon Solenoid) multi-purpose detectors fulfil the requirements needed to find the Higgs boson in the various final states. Beside an excellent detector performance, outstanding trigger systems will reduce the huge event rate of 40 MHz (this rate corresponds to nearly 10$^9$ interactions per second) to an acceptable rate of about 100 Hz.

\section{Standard Model Searches}

During the last years, a lot of progress have been made in understanding Higgs boson production and decays from a theoretical point of view. For example, the production cross section of gluon fusion has been calculated at NNLO (Next-to-Next-to-Leading Order). Calculations at higher orders yield a better prediction of the cross section. The kinematic properties of the events can change as well, which result in a different selection efficiency and therefore in a different number of events collected in the end.

\begin{figure}
  \centering
  \hspace*{-4mm}\epsfig{figure=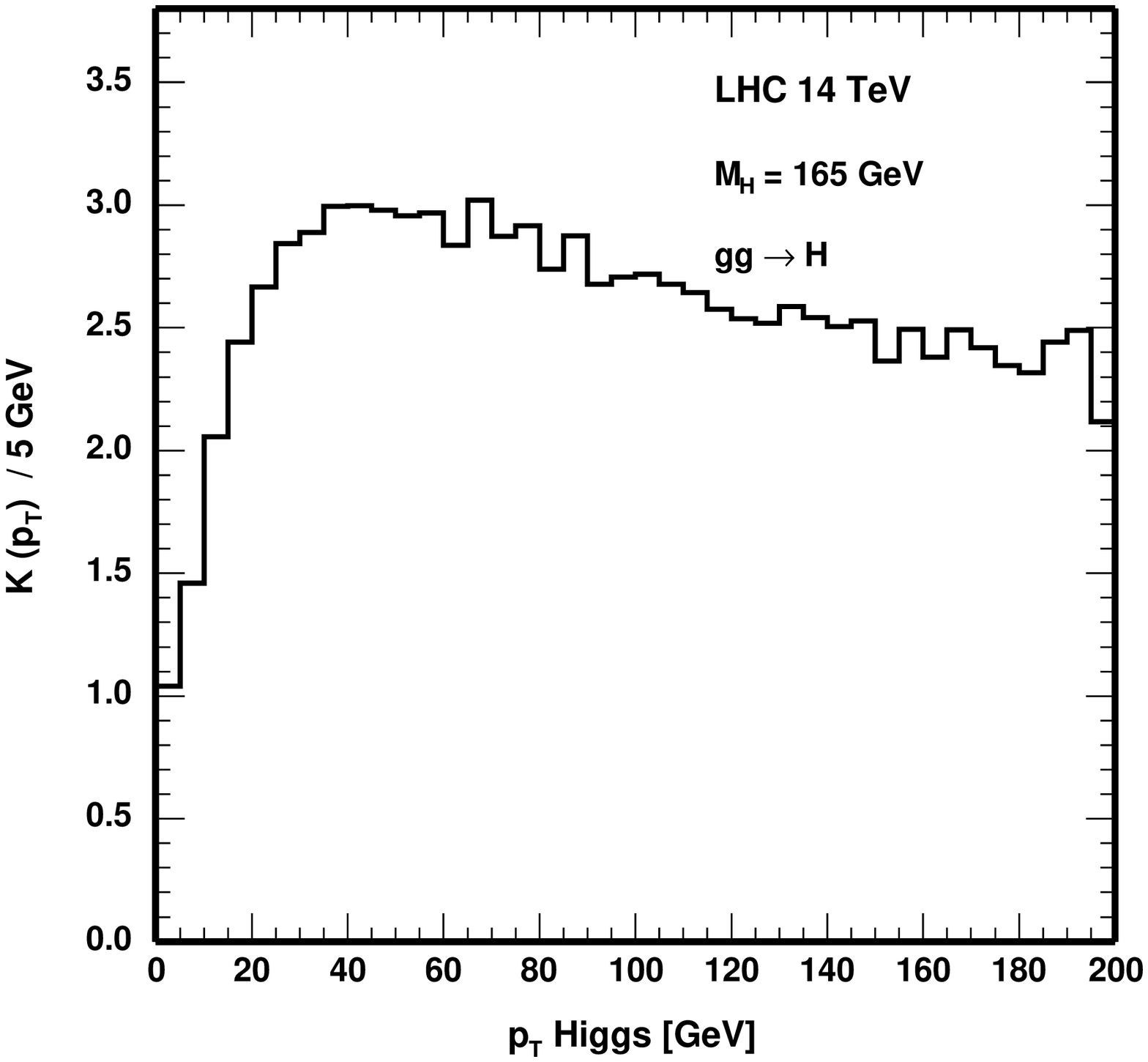,width=0.41\textwidth}
  \hspace*{6mm}\epsfig{figure=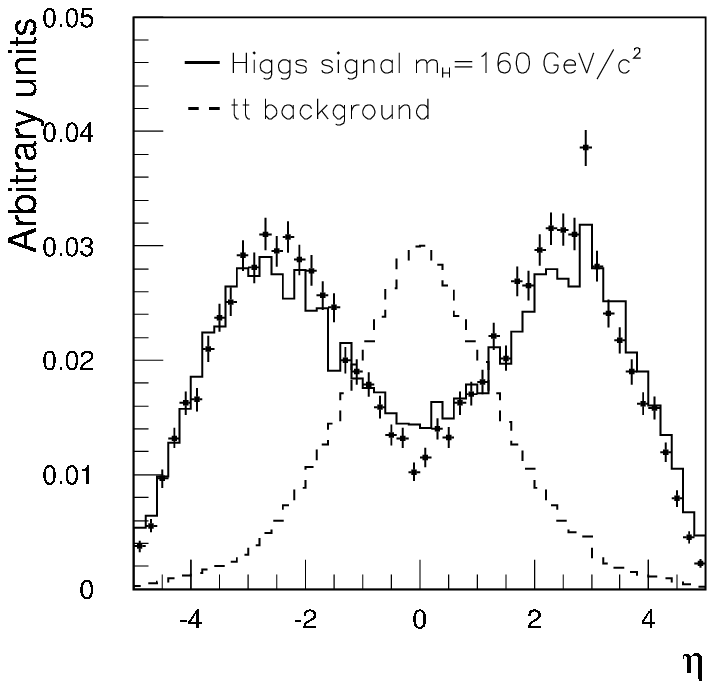,width=0.38\textwidth}
  \put(-345,145){\scalebox{1.1}[1.1]{a)}}
  \put(-130,120){\scalebox{1.1}[1.1]{b)}}
  \caption{a) $k$ factor distribution as a function of the transverse momentum of the Higgs boson. b) ${\rm \eta}$ distributions for jets coming from vector boson fusion (histograms: parton level; dots: after reconstruction) or from ${\rm t\bar{t}}$ events.
  \label{fig:eff_k_fac}}
\end{figure}

In Ref.~\cite{ref:k_fact}, the channel ${\rm gg \rightarrow h \rightarrow W^+ W^- \rightarrow \ell^+ \nu \ell^- \bar{\nu}}$ is examined. A jet veto (no jet with $p{\rm _T(jet) > 20\ GeV/}c$ in the detector acceptance) reduces the backgrounds coming from events with top quarks. Since higher-order effects include additional gluon radiation, the $k$ factor is affected by this cut.
The differential $k$ factor is defined as $k{\rm (\xi) = \frac{\partial\sigma_{NNLO}(\xi)\ /\ \partial\xi}{\partial\sigma_{MC - LO}(\xi)\ /\ \partial\xi}}$,
where ${\rm \xi}$ could be any kinematic variable, and ${\rm \sigma_{MC - LO}(\xi)}$ is the leading order cross section obtained with an event generator. The use of an event generator is necessary in order to have a realistic final state with fragmented particles and underlying event. The $k$ factor is shown in Fig.~\ref{fig:eff_k_fac}a as a function of the transverse momentum of the Higgs boson. This variable is accessible in both NNLO calculation and event generator. The Higgs boson transverse momentum is closely connected to the jet veto cut because a jet would be balanced with the Higgs boson in the transverse plane. The $k$ factor from Fig.~\ref{fig:eff_k_fac}a is used to reweight individual events with reconstructed jets coming from a LO event generator. This technique can be applied to other processes equivalently, for example to the ${\rm W^+ W^-}$ background in this channel.

The result is an effective $k$ factor of 2.04 for $m{\rm _h = 165\ GeV/}c^2$ which is  about 15\% lower than the inclusive $k$ factor for the same mass. Effective $k$ factors not only help to predict the discovery potential more accurately, but also allow a better cut optimization.

Another development in Standard Model Higgs physics is to exploit the kinematic features of Higgs bosons produced in vector boson fusion~\cite{ref:atl_vbf}$^,\ $\cite{ref:cms_vbf}. In this process, the Higgs boson is produced together with two energetic jets. As shown in Fig.~\ref{fig:eff_k_fac}b, one jet is produced typically in the forward direction, and the other one in the backward direction. Jets from background processes are either produced central or both in the forward (backward) direction. Several cuts reduce the backgrounds effectively: high energies, high difference in ${\rm \eta}$, and high invariant mass of the two jets. In addition, the absence of colour connection between the two jets produced in vector boson fusion leads to a very low hadronic activity and can be exploited by a veto on additional jets in the ${\rm \eta}$ region between the two jets mentioned before.

\begin{figure}[b]
  \centering
  \epsfig{figure=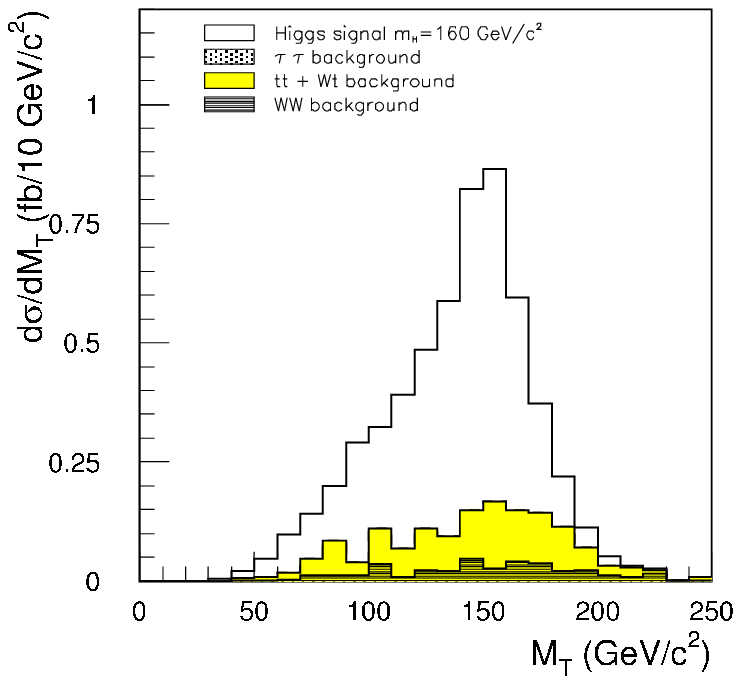,width=0.34\textwidth}
  \epsfig{figure=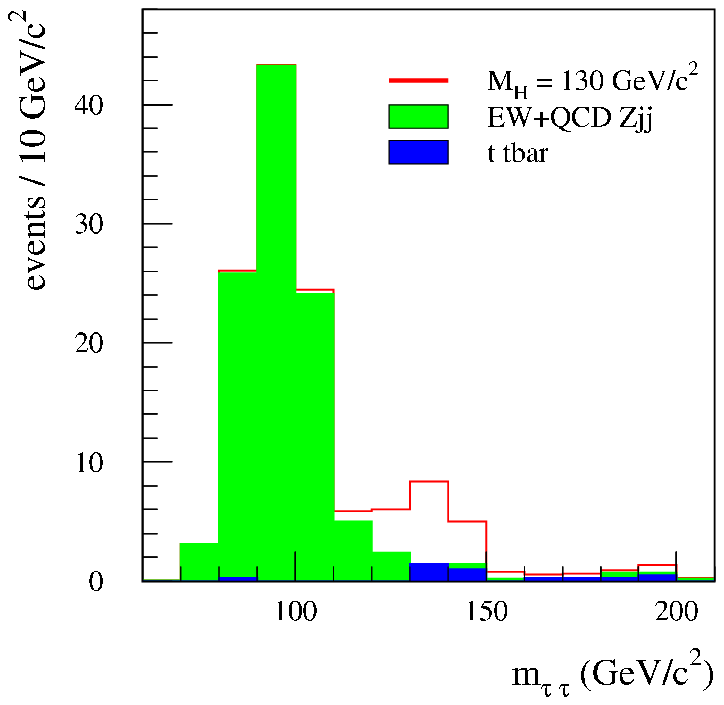,width=0.31\textwidth}
  \hspace*{2mm}\epsfig{figure=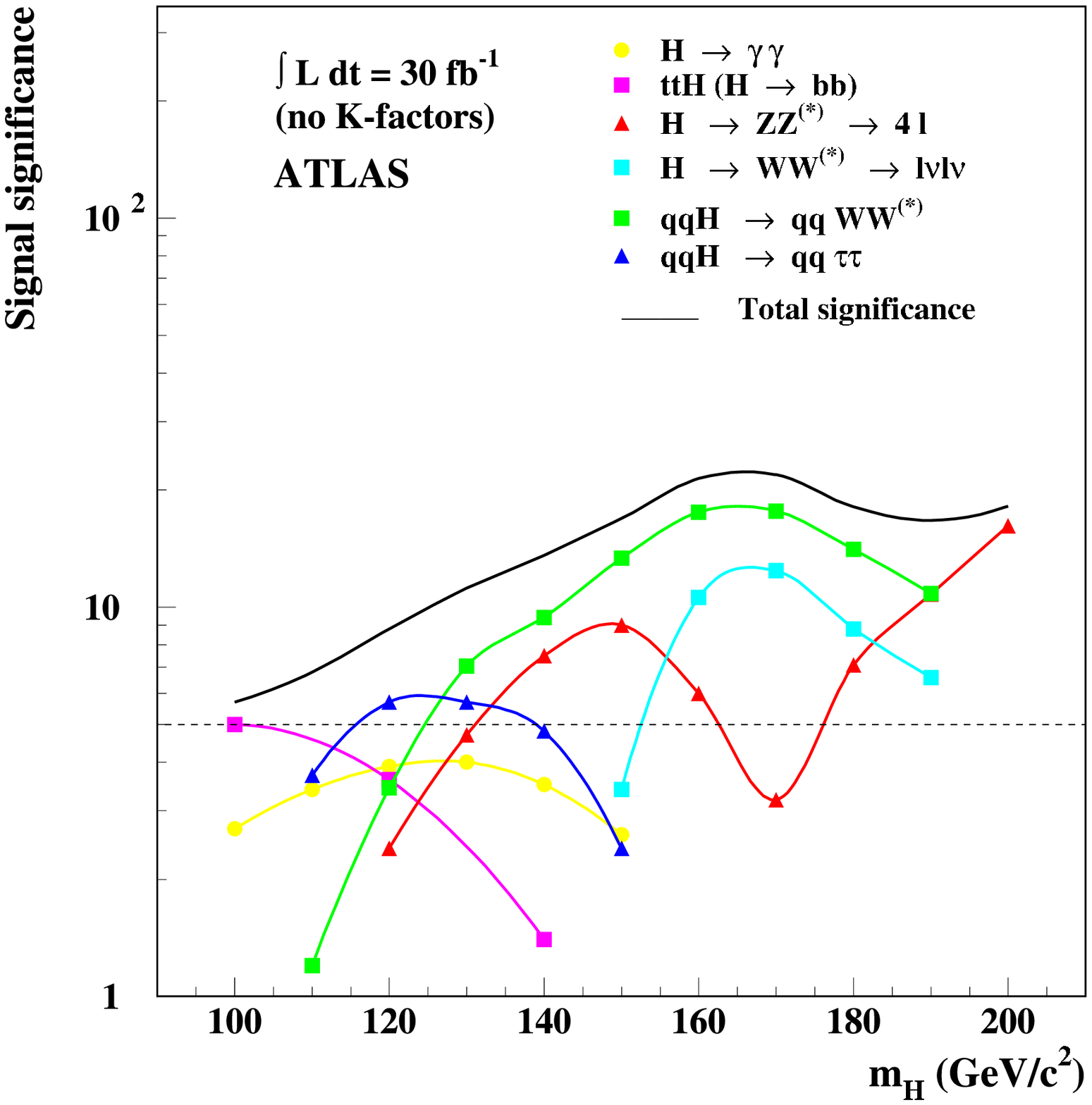,width=0.31\textwidth}
  \put(-419,130){\scalebox{1.1}[1.1]{a)}}
  \put(-260,130){\scalebox{1.1}[1.1]{b)}}
  \put(-118,130){\scalebox{1.1}[1.1]{c)}}
  \caption{Examples of vector boson fusion: a) Transverse mass distribution for ${\rm h \rightarrow W^+ W^- \rightarrow \ell^+ \nu \ell^- \bar{\nu}}$. b) Signal and background rates for ${\rm h \rightarrow  \tau^+ \tau^- \rightarrow \ell^\pm j + X}$. c) Discovery potential for the Standard Model Higgs boson.
  \label{fig:atl_vbf}}
\end{figure}

\begin{figure}
  \centering
  \vspace*{-5mm}
  \hspace*{-4mm}\epsfig{figure=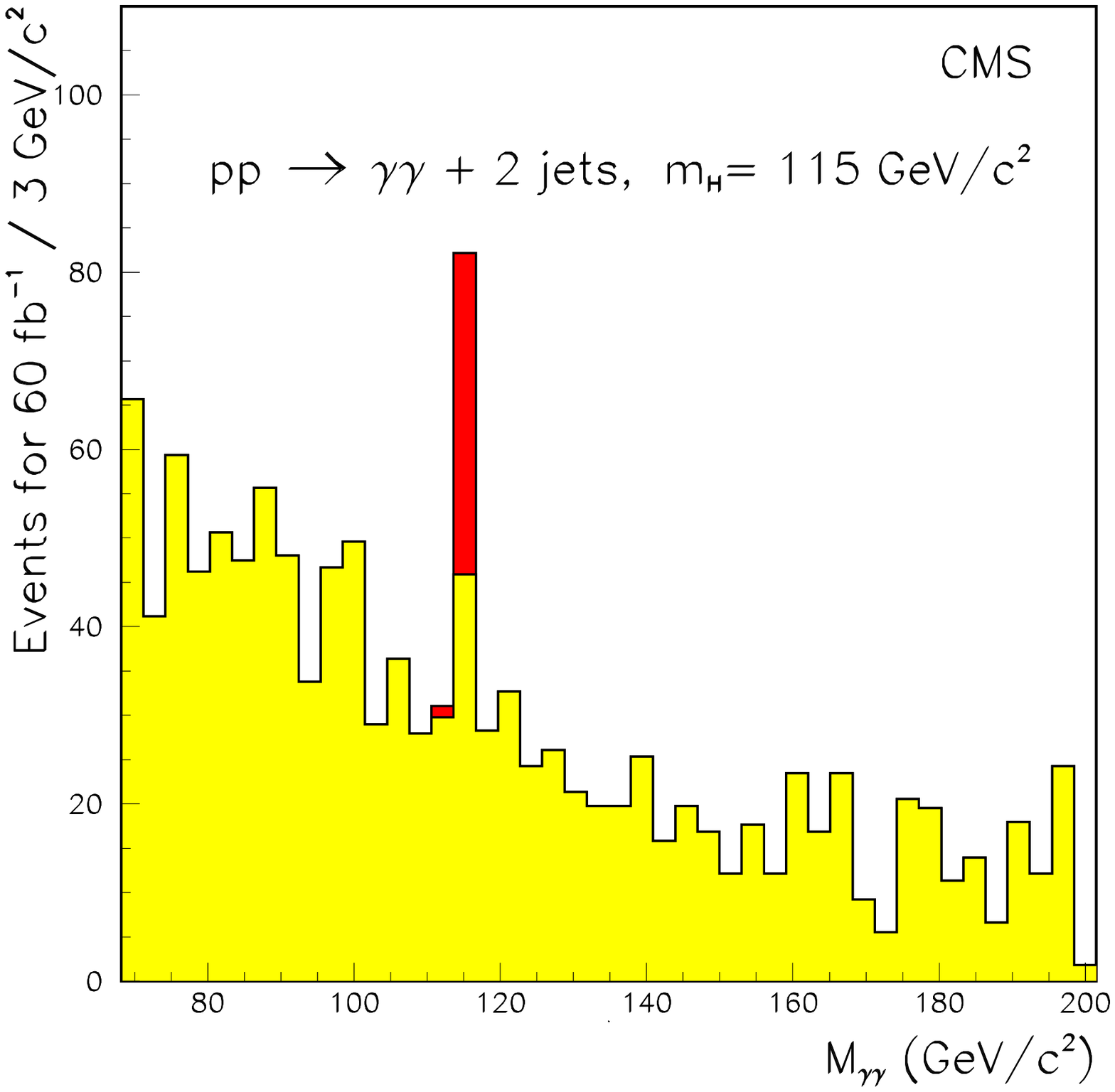,width=0.40\textwidth}
  \hspace*{2mm}\epsfig{figure=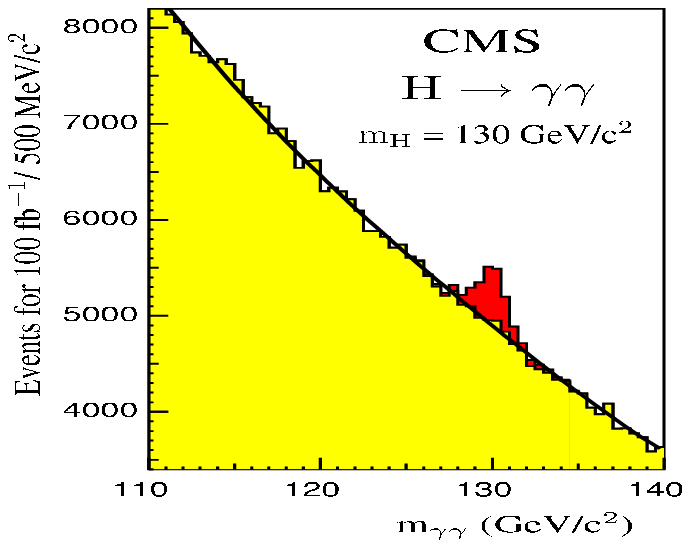,width=0.467\textwidth}
  \put(-380,150){\scalebox{1.1}[1.1]{a)}}
  \put(-140,150){\scalebox{1.1}[1.1]{b)}}
  \caption{${\rm \gamma \gamma}$ mass distributions for ${\rm h \rightarrow \gamma \gamma}$ decay channels: a) vector boson fusion production. b) inclusive production. The signal (red/dark) to background (yellow/light) ratio is increased by the additional jet requirements.
  \label{fig:cms_vbf}}
  \vspace*{-3mm}
\end{figure}

The actual values for the cuts depend on the Higgs boson decay mode. Clean final states require relatively soft cuts. Channels with high backgrounds need stronger jet-tagging requirements. Figures~\ref{fig:atl_vbf} and \ref{fig:cms_vbf} give an idea of the benefit of the additional jet requirements. In the decay mode ${\rm h \rightarrow W^+ W^- \rightarrow \ell^+ \nu \ell^- \bar{\nu}}$, the background is removed almost completely. A large signal-to-background ratio is particularly important in this channel and could be even larger, because the systematic uncertainties are quite high. A background uncertainty of 10\% has been assumed, because no mass peak can be reconstructed in this decay mode.

The decay ${\rm h \rightarrow  \tau^+ \tau^-}$ with one ${\rm \tau^\pm}$ decaying hadronically and the other leptonically can be detected as well. Under the assumption that the neutrinos of the ${\rm \tau^\pm}$ decays are collinear to the visible decay products, the mass of the Higgs boson can be reconstructed from the ${\rm \tau^\pm}$ jet, the lepton and the missing transverse energy. Without this special identification of the two jets, the Standard Model Higgs boson is not visible in the ${\rm h \rightarrow  \tau^+ \tau^-}$ decay mode.

As a last example, the ${\rm h \rightarrow \gamma \gamma}$ channel gains a lot from the typical jet topology of the vector boson fusion process, as is illustrated in Fig.~\ref{fig:cms_vbf}a. The comparison with Fig.~\ref{fig:cms_vbf}b shows a clear improvement of the signal-to-background ratio. In general, the signal identification by exploiting the features of vector boson fusion helps to control the various backgrounds better. These additional channels improve the discovery potential of the Standard Model Higgs boson, as shown in Fig.~\ref{fig:atl_vbf}c, and are also useful to measure the couplings of the Higgs boson to other particles, in particular to vector bosons.

\section{Parameter Measurements}

\begin{figure}
  \centering
  \hspace*{-4mm}\epsfig{figure=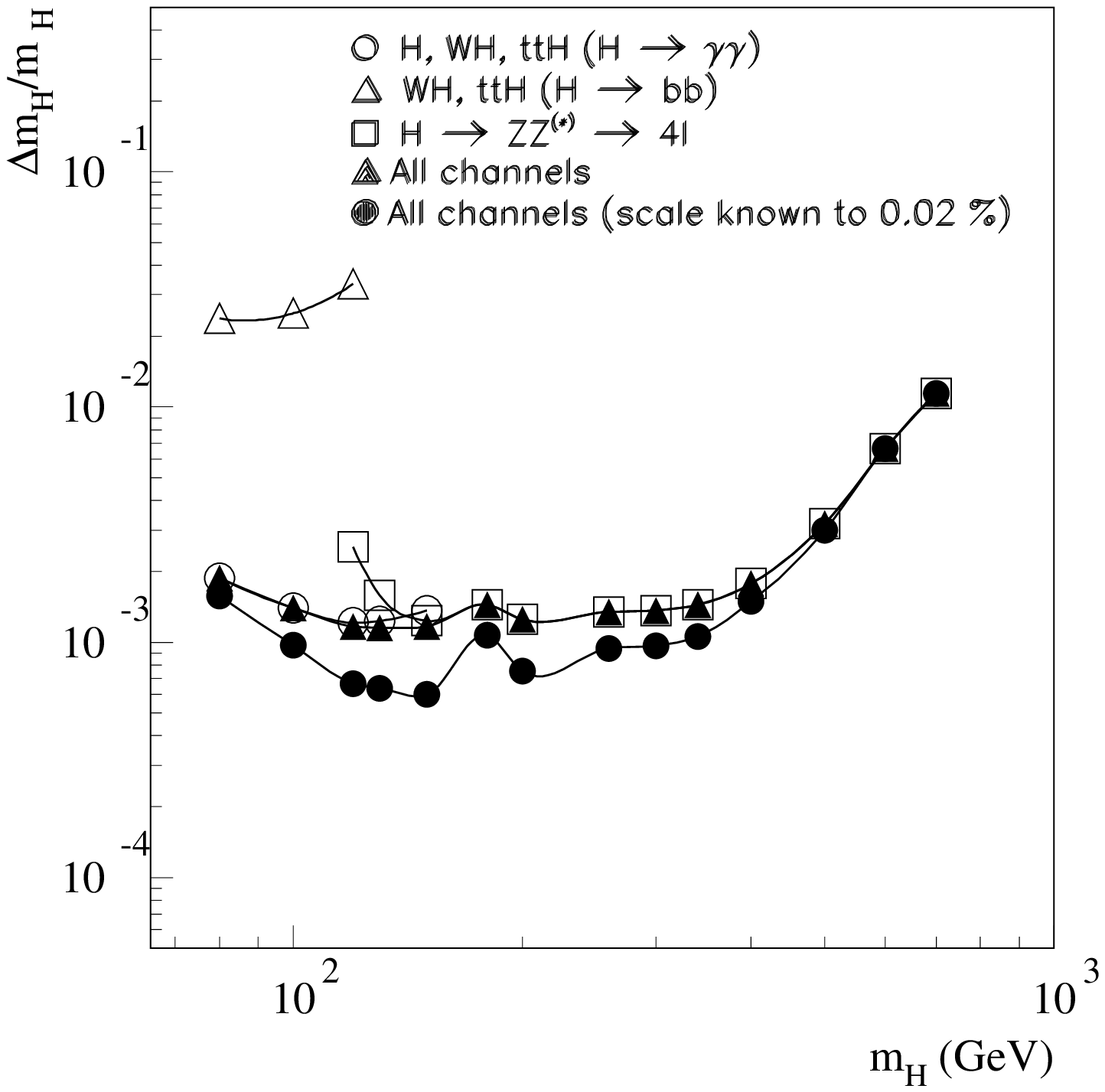,width=0.386\textwidth}
  \hspace*{6mm}\epsfig{figure=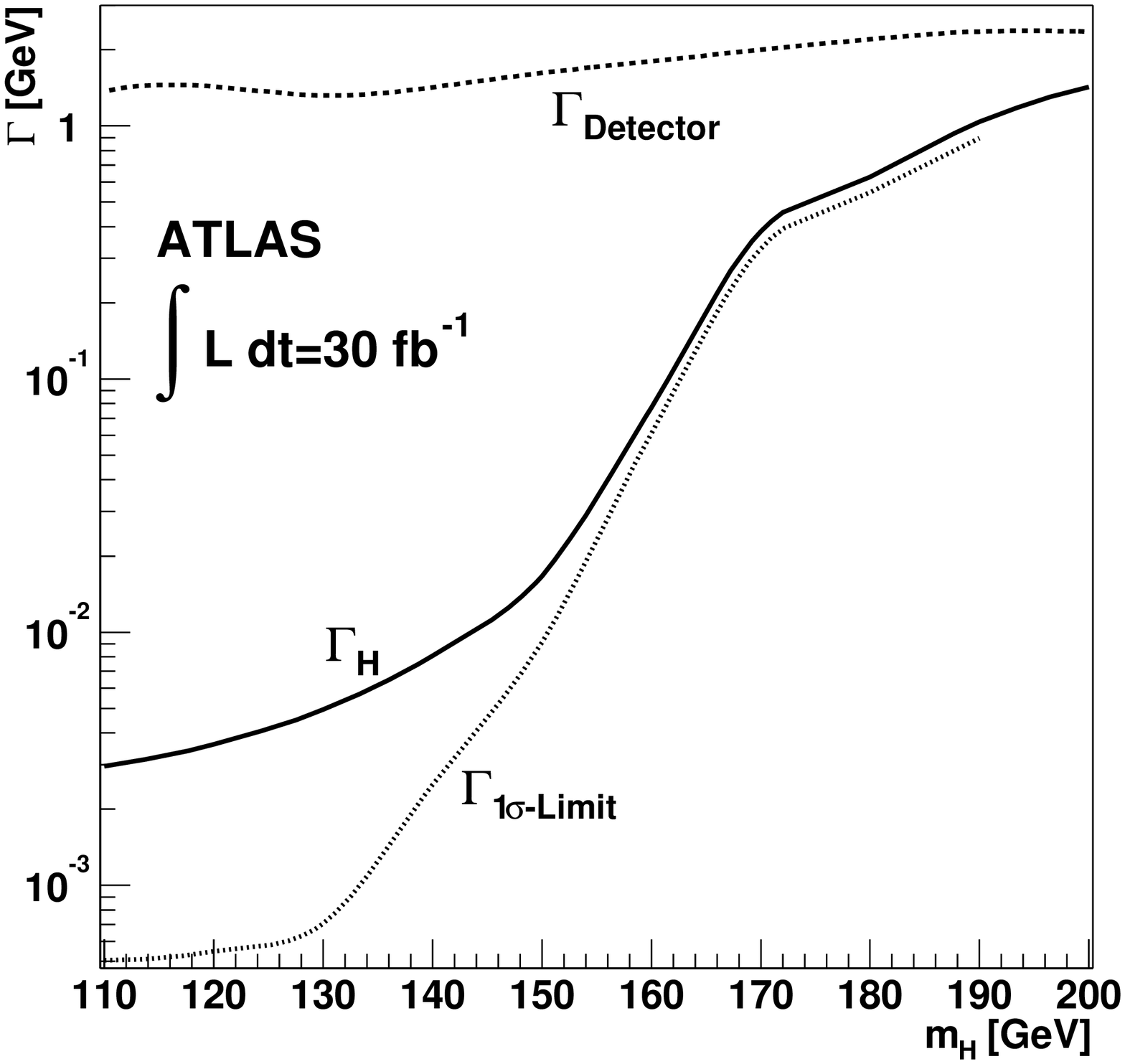,width=0.40\textwidth}
  \put(-350,160){\scalebox{1.1}[1.1]{a)}}
  \put(-158,160){\scalebox{1.1}[1.1]{b)}}
  \caption{a) Precision of the Higgs boson mass measurement for ${\rm L_{int} = 300\ fb^{-1}}$. An uncertainty of 0.1\% is assumed for the electromagnetic. energy scale. b) Total width of the Standard Model Higgs boson, lower limit obtained from indirect determination of the width, and upper limit coming from direct measurements.
  \label{fig:mass_width}}
\end{figure}

After a discovery, it is important to find out whether the particle is indeed a Higgs boson and to understand the underlying model. A way to address these questions is to measure as many properties of the Higgs boson as possible. Since most of the analyses are based on the search for a mass bump, the measurement of the Higgs boson mass comes almost for free~\cite{ref:atl_tdr}. Channels for which the invariant mass cannot be reconstructed still provide kinematic distributions (such as that of the transverse mass) sensitive to the mass of the Higgs boson. As shown in Fig.~\ref{fig:mass_width}a, the precision of the mass measurement is better than 1\% over the whole mass range studied. The width can also be measured directly from a fit to the mass peak, if the natural width of the Higgs boson is comparable to or larger than the corresponding detector resolution. In the Standard Model, this is the case for $m{\rm _h > 200\ GeV/}c^2$, as illustrated in Fig.~\ref{fig:mass_width}b. For smaller masses, the width must be determined indirectly~\cite{ref:couplings}. An updated study which includes a few theoretical assumptions has been presented at this conference~\cite{ref:g_talk}: the width and many couplings can be measured at the level of about 30\%. This may be enough to check if the Higgs boson couples to the mass of other particles and to find major deviations from a scenario under investigation.\\\

\begin{figure}[b]
  \centering
  \epsfig{figure=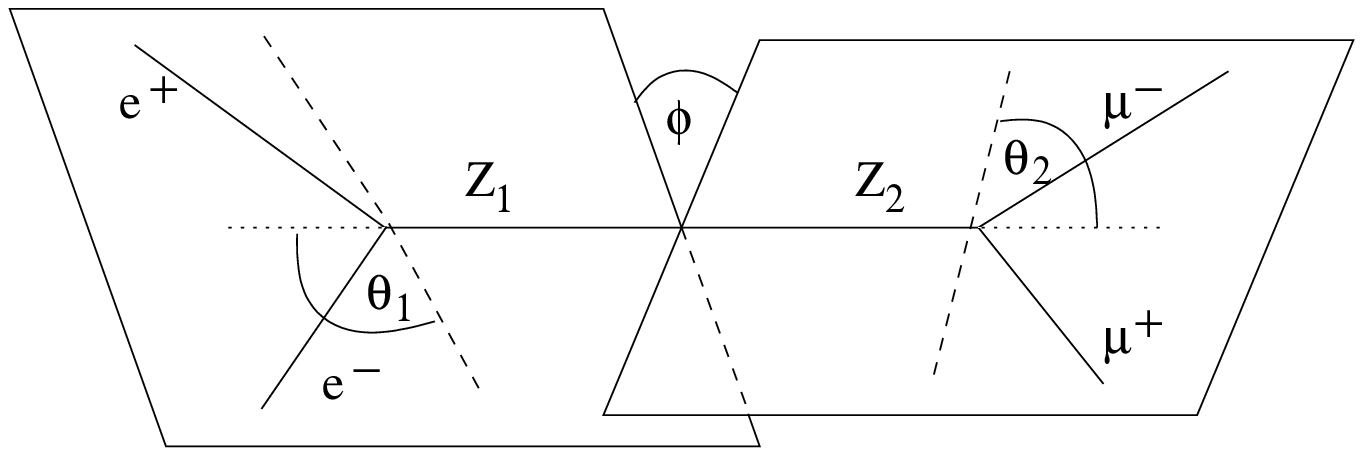,width=0.55\textwidth}
  \hspace*{2mm}\epsfig{figure=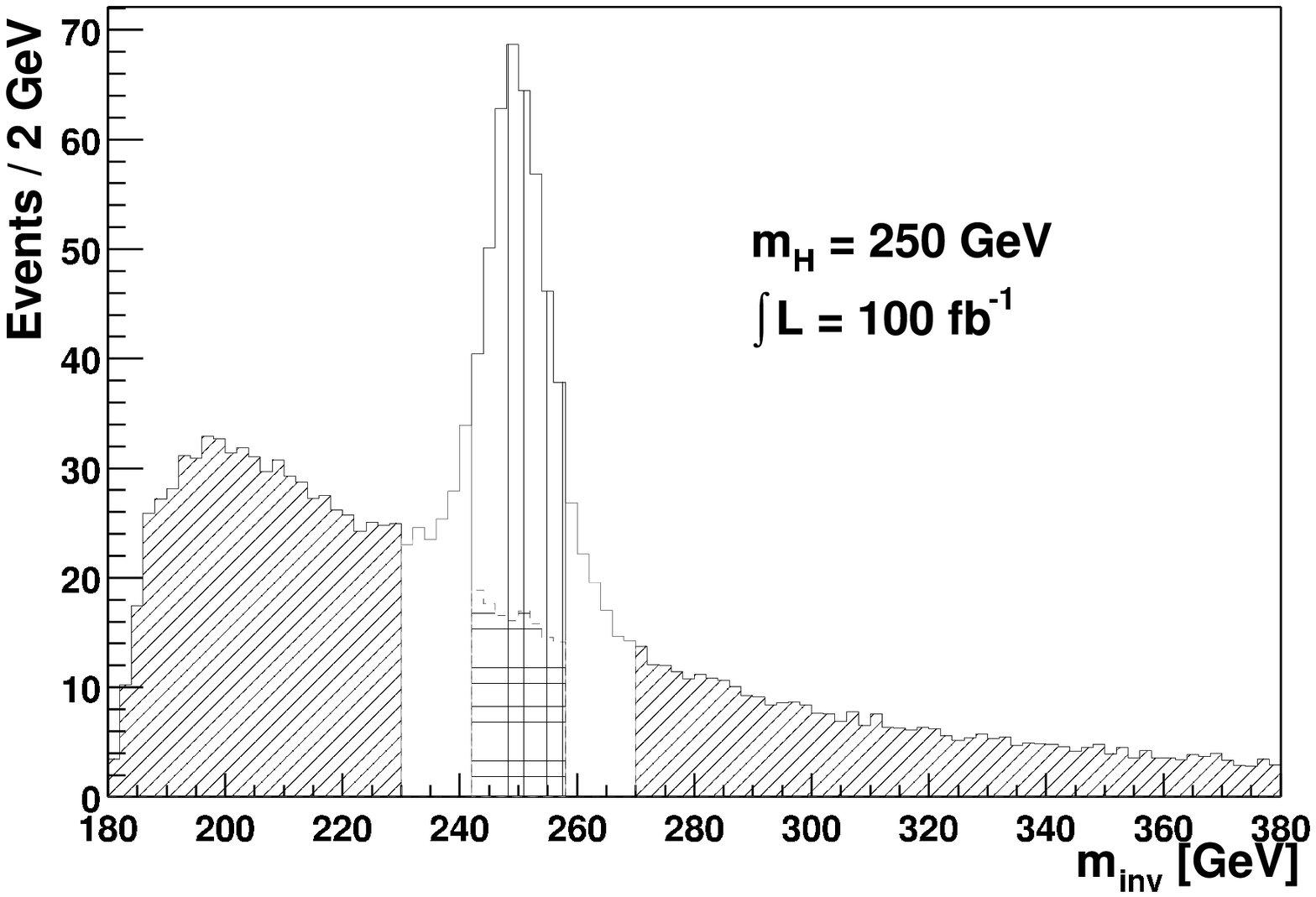,width=0.35\textwidth}
  \put(-425,75){\scalebox{1.1}[1.1]{a)}}
  \put(-138,85){\scalebox{1.1}[1.1]{b)}}
  \caption{a) Definitions of decay plane angle ${\rm \phi}$ and the polarization angles ${\rm \theta_{1,2}}$ in ${\rm h \rightarrow Z Z}$ events. b) Four lepton invariant mass distribution; hatched regions are signal and background regions used for sideband subtraction.
  \label{fig:spin_cp}}
\end{figure}

Other important properties of the Higgs boson are its spin and its CP eigenvalue. The Standard Model Higgs boson is a CP-even scalar. In the MSSM (the Minimal Supersymmetric extension of the Standard Model), there are two CP-even, one CP-odd, and two charged Higgs bosons. The charged Higgs bosons can be identified by their different decay modes. Spin and CP eigenvalues can be measured simultaneously in the channel ${\rm h \rightarrow Z Z \rightarrow \ell^+ \ell^- \ell^+ \ell^-}$ over a wide mass range, ${\rm 200\ GeV/}c^2 < m{\rm _h < 400\ GeV/}c^2$ is investigated in Ref.~\cite{ref:spin_cp}. In this process, three angles ${\rm \phi}$ and ${\rm \theta_{1,2}}$ are defined, as sketched in Fig.~\ref{fig:spin_cp}. They are sensitive to spin and CP of the Higgs boson. After the straightforward selection of four isolated leptons, and reconstruction of two Z boson and one Higgs boson mass peaks, the background is subtracted statistically using events outside the signal region, as illustrated in Fig.~\ref{fig:spin_cp}b. From fits to the angular distributions, three parameters ($\alpha$, $\beta$ and $R$) are extracted. The values of these parameters can be compared to theoretical predictions of spin = 0,1 and CP = $\pm$ 1 hypotheses. With an integrated luminosity of 100 ${\rm fb^{-1}}$, spin equal to unit can be ruled out at 95\% confidence level. A CP-odd Higgs boson can be discriminated with less integrated luminosity.\\\

The vacuum expectation value $v$ of the Higgs field could be accessed through the trilinear Higgs coupling, proportional to $ 1/v^2$. In principle, it is possible to get hold of the trilinear Higgs coupling by measuring the cross sections of the following processes ${\rm h \rightarrow h h \rightarrow b\bar{b}\gamma \gamma}$ or ${\rm h \rightarrow h h \rightarrow b\bar{b}\mu^+ \mu^-}$ at small Higgs boson mass and with the decay mode ${\rm h \rightarrow W^+ W^-}$ at larger masses. This turns out to be very challenging~\cite{ref:tri_hig}.

Things are a bit easier when it comes to measure the ratio of the vacuum expectation values ${\rm \tan\beta =\ } v_2/v_1$ in two-Higgs-doublet models. In the following, two examples~\cite{ref:cms_tb}$^,\ $\cite{ref:atl_tb} of the MSSM are discussed. In both cases, the determination of ${\rm \tan\beta}$ is based on the measurement of the event rate or cross section times branching ratio, respectively.
\begin{figure}
  \centering
  \vspace*{-5mm}
  \hspace*{0mm}\epsfig{figure=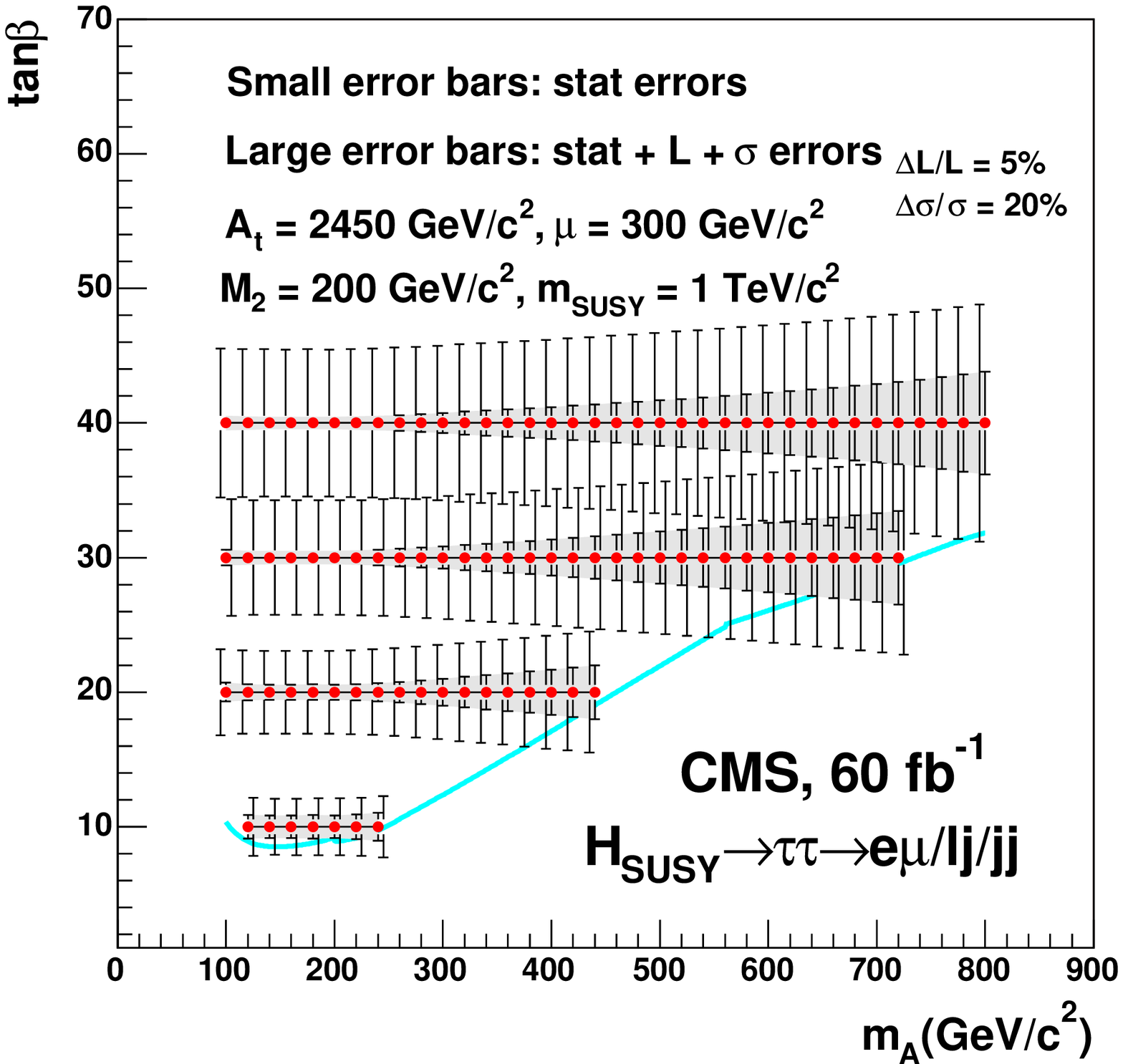,width=0.32\textwidth}
  \hspace*{0mm}\epsfig{figure=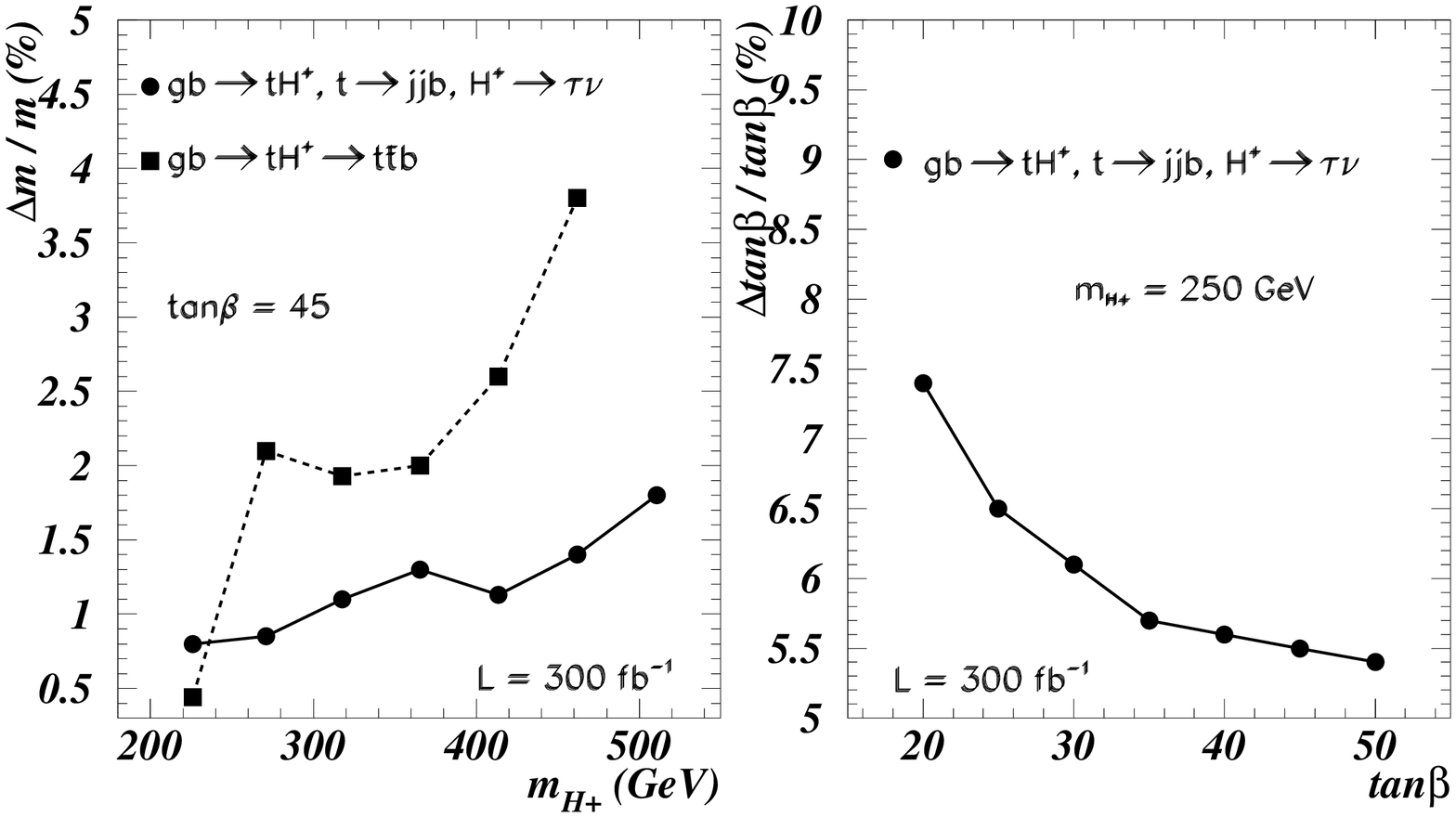,width=0.55\textwidth}
  \put(-280,112){\scalebox{1.1}[1.1]{a)}}
  \put(-145,112){\scalebox{1.1}[1.1]{b)}}
  \put(-33,112){\scalebox{1.1}[1.1]{c)}}
  \caption{a) ${\rm \tan\beta}$ measurement with ${\rm b\bar{b}h/H/A \rightarrow b\bar{b}\tau^+\tau^-}$ events as a function of $m{\rm _A}$ and ${\rm \tan\beta}$. Precision of b) $m{\rm _{H^\pm}}$ and c) ${\rm \tan\beta}$ for ${\rm gb \rightarrow tH^\pm}$ charged Higgs boson production with the decay $H^\pm \rightarrow \tau^\pm \nu$.
  \label{fig:tan_beta}}
\end{figure}

In the neutral Higgs sector, information on ${\rm \tan\beta}$ can be obtained by measuring the event rate in the channel ${\rm b\bar{b}h/H/A \rightarrow b\bar{b}\tau^+\tau^-}$ with ${\rm \tau^\pm}$ either decaying hadronically or leptonically. The leading terms of the production cross section are proportional to ${\rm \tan^2 \beta}$. The analysis is based on the requirement of exactly one b jet and the reconstruction of the Higgs boson mass from the  ${\rm \tau^+\tau^-}$ pair. While the identification of one b jet reduces the background considerably, a second b tag does not increase the statistical significance. The result for the ${\rm \tan\beta}$ measurement is shown in Fig.~\ref{fig:tan_beta}a which includes a theoretical uncertainty of 20\% on the production cross section. This large uncertainty can be reduced to the level of 10\% when the presence of two b quarks is required which might improve the ${\rm \tan\beta}$ measurement in some cases. Other uncertainties, included in this study, are 3\% for the Higgs branching ratio and 5\% for the luminosity.

A similar analysis is performed for the charged Higgs boson. Here the production cross section (here ${\rm gb \rightarrow tH^\pm}$) is sensitive to ${\rm \tan\beta}$, too. The decay mode ${\rm H^\pm \rightarrow \tau^\pm \nu}$ gives the best coverage of the MSSM parameter space. In this channel, only the reconstruction of the transverse mass is possible. As can be seen in Figs.~\ref{fig:tan_beta}b and~\ref{fig:tan_beta}c, this analysis leads to a mass measurement with a precision of about 1\% and ${\rm \tan\beta}$ can be measured with a precision better than 10\% including systematic uncertainties. It has to be pointed out that both ${\rm \tan\beta}$ measurements are only correct, if there is no undetected decay mode with sizeable branching ratio and if the other model parameters are known. A cross check is possible by comparing both measurements, based on independent decay modes.

\section{Exotic Scenarios}
Exotic scenarios are defined here as cases that are not predicted by the Standard Model. Two types of scenarios are discussed: a) heavy particles decaying to Higgs bosons and b) decays of Higgs bosons to neutralinos, which leads to partly invisible final states.\\\

\begin{figure}
  \centering
  \vspace*{-4mm}
  \hspace*{-5mm}\epsfig{figure=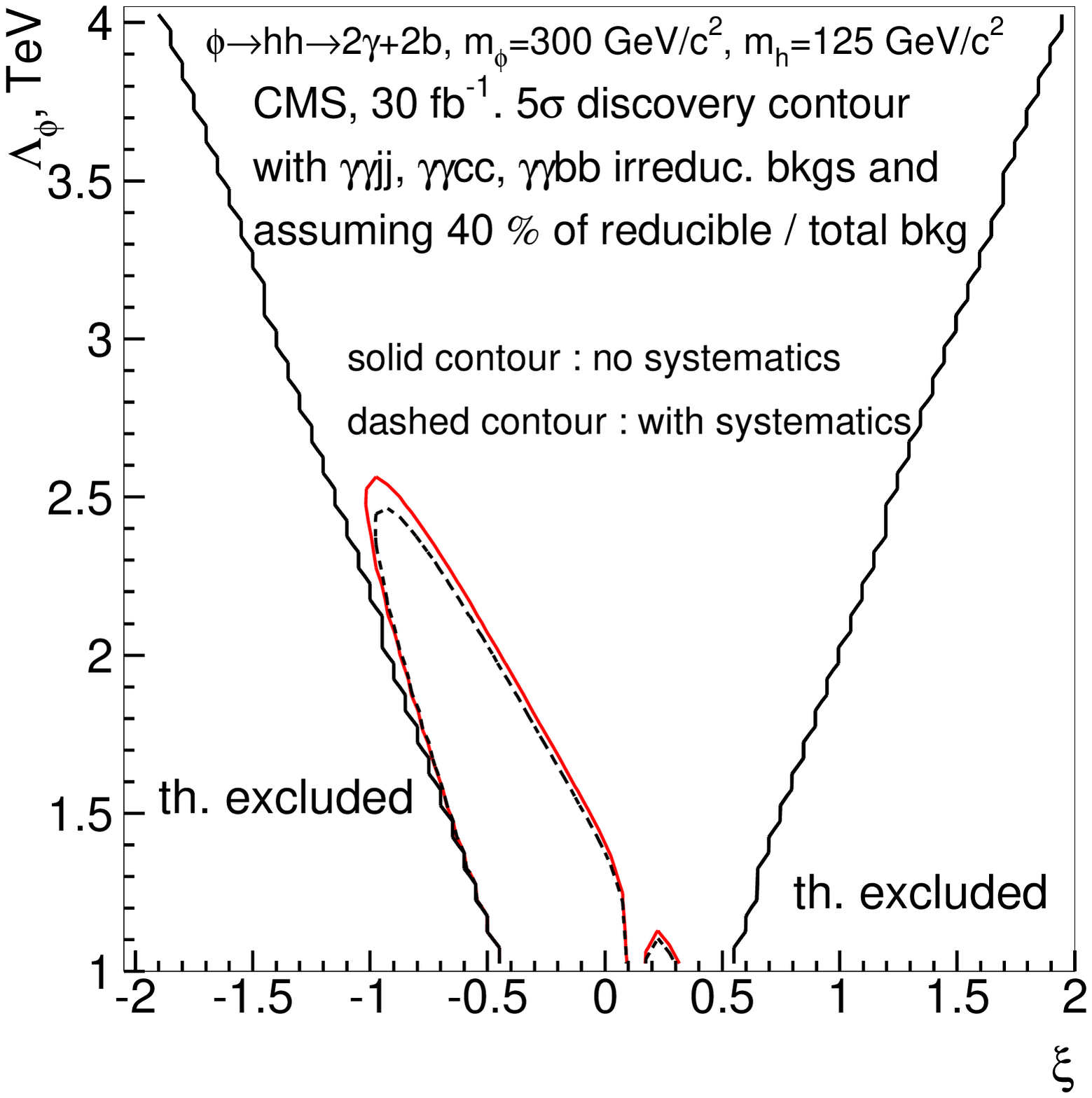,width=0.33\textwidth}
  \hspace*{2mm}\epsfig{figure=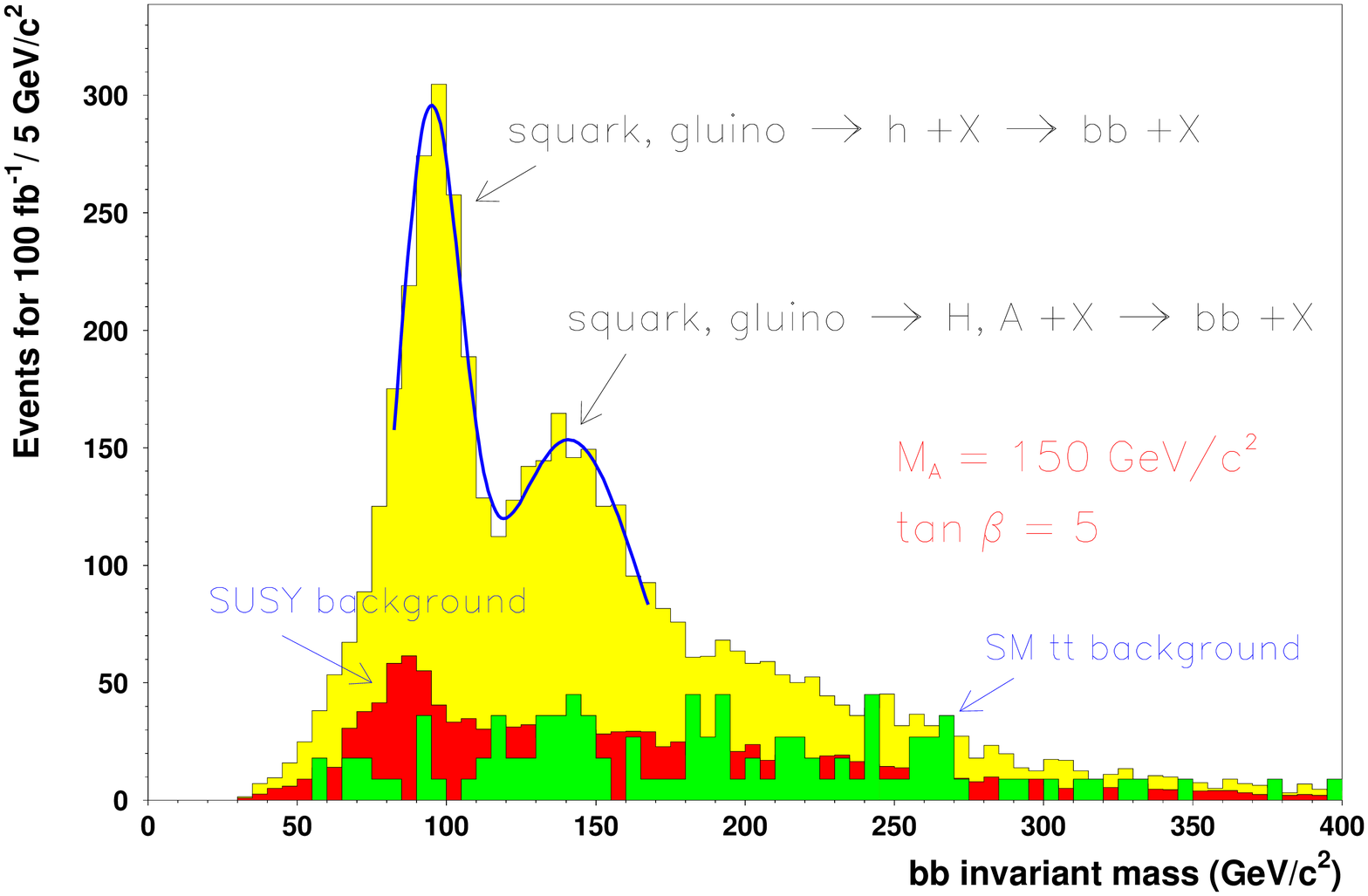,width=0.505\textwidth}
  \put(-368,110){\scalebox{1.1}[1.1]{a)}}
  \put(-200,128){\scalebox{1.1}[1.1]{b)}}
  \vspace*{-0.5mm}
  \caption{a) 5 ${\rm \sigma}$-discovery contour for a 300 ${\rm GeV/}c^2$ radion decaying to a pair of Higgs bosons into the ${\rm \gamma \gamma b\bar{b}}$ final state. b) Signal and background for Higgs bosons produced in supersymmetric cascades with ${\rm h/H/A \rightarrow b\bar{b}}$.
  \label{fig:exo_decays}}
\end{figure}

Models with extra dimensions predict additional particles as the radion ${\rm \phi}$ which can, if heavy enough, decay to a pair of Higgs bosons. Beside the masses $m{\rm _\phi}$ and $m{\rm _h}$, the main parameters in models with one extra dimension are a dimensionless parameter ${\rm \xi}$ and the vacuum expectation value of the radion field ${\rm \Lambda_\phi}$. In Ref.~\cite{ref:radion}, a radion mass of 300 ${\rm GeV/}c^2$ and a Higgs boson mass of ${\rm 125\ GeV/}c^2$ are assumed. In this example, the channel ${\rm \phi \rightarrow h h \rightarrow \gamma \gamma b\bar{b}}$ is considered. In this process, three resonances can be reconstructed: a mass peak from ${\rm h \rightarrow \gamma \gamma}$, a second peak from ${\rm h \rightarrow b\bar{b}}$, and finally, the mass of the radion reconstructed from the whole ${\rm \gamma \gamma b\bar{b}}$ final state. As a result, the 5 ${\rm \sigma}$-discovery potential is shown in Fig.~\ref{fig:exo_decays}a as a function of ${\rm \xi}$ and ${\rm \Lambda_\phi}$.

Another interesting possibility to produce Higgs boson is the cascade decays in supersymmetric models~\cite{ref:cms_vbf}. A supersymmetric cascade typically starts with the production of a pair of strongly interacting particles, for example a squark and a gluino. Then a chain of consecutive decays starts. One possible way is ${\rm \tilde{q} \rightarrow q \chi^0_2 \rightarrow q \chi^0_1 h\rightarrow q \chi^0_1 b\bar{b}}$ and ${\rm \tilde{g} \rightarrow \tilde{t}t}$ with consecutive decays. There are many possibilities to build such final states, which typically consist of many b jets, missing transverse energy coming from ${\rm \chi^0_1}$s or ${\rm \nu}$s , and other particles. Since there are many final sates possible and the final states are quite complex, a generic event selection is the first choice. Events with large missing transverse energy and many jets, tagged as b jets, are selected. The pair of the two closest b jets is chosen to reconstruct the Higgs boson invariant mass. The result of this reconstruction is shown in Fig.~\ref{fig:exo_decays}b. In this case, the mass spectrum shows one peak coming from the lightest MSSM Higgs boson ${\rm h}$ and a second peak coming from ${\rm H}$ and ${\rm A}$. Because of the extraordinary number of particles in the final state, the background is low in general, but combinatorial background has to be considered. The underlying model for this example is the MSSM, but similar scenarios are possible in any kind of supersymmetric scenario.\\\

\begin{figure}[hb]
  \centering
  \vspace*{-7mm}
  \hspace*{-5mm}\epsfig{figure=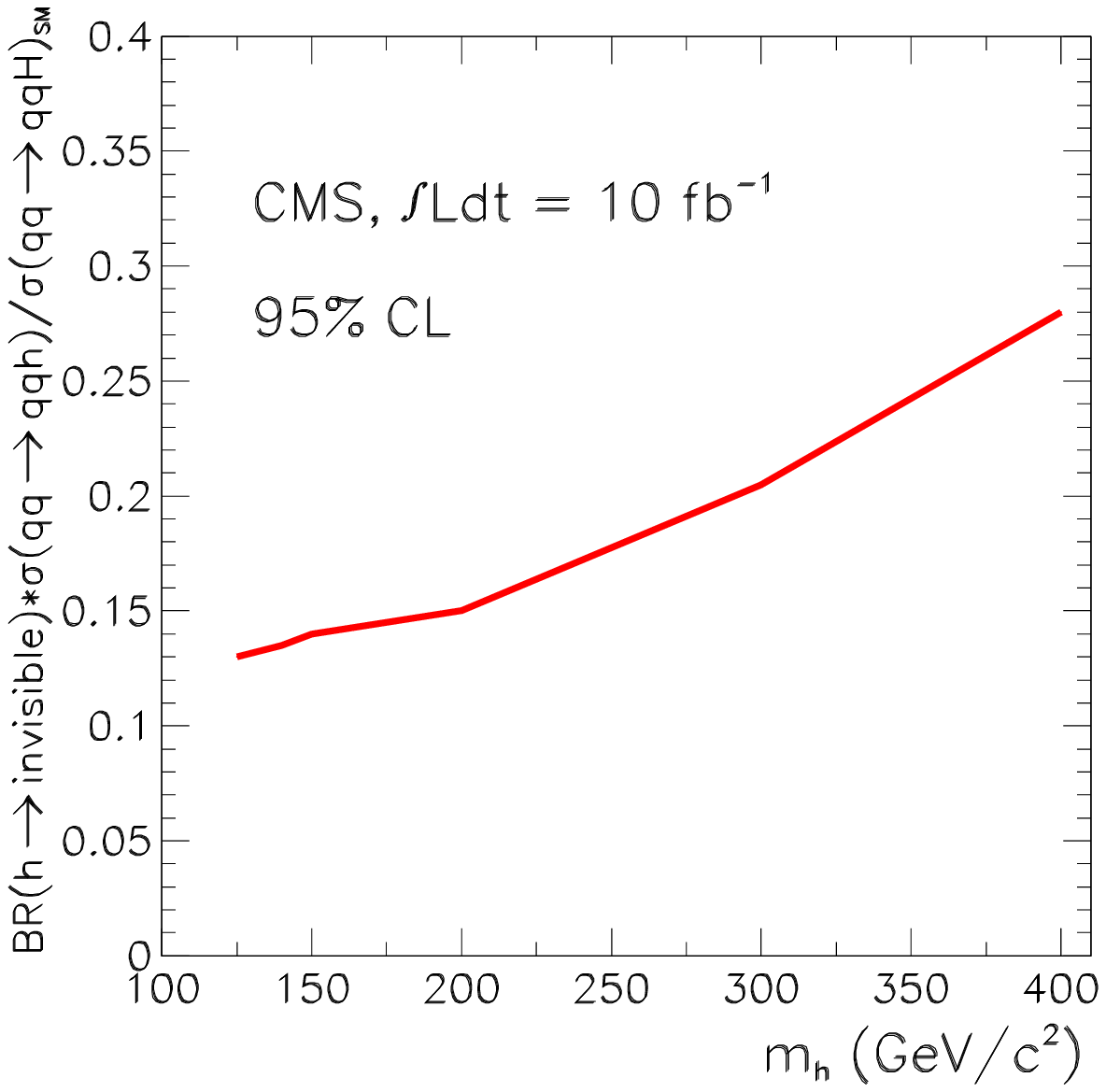,width=0.37\textwidth}
  \hspace*{2mm}\epsfig{figure=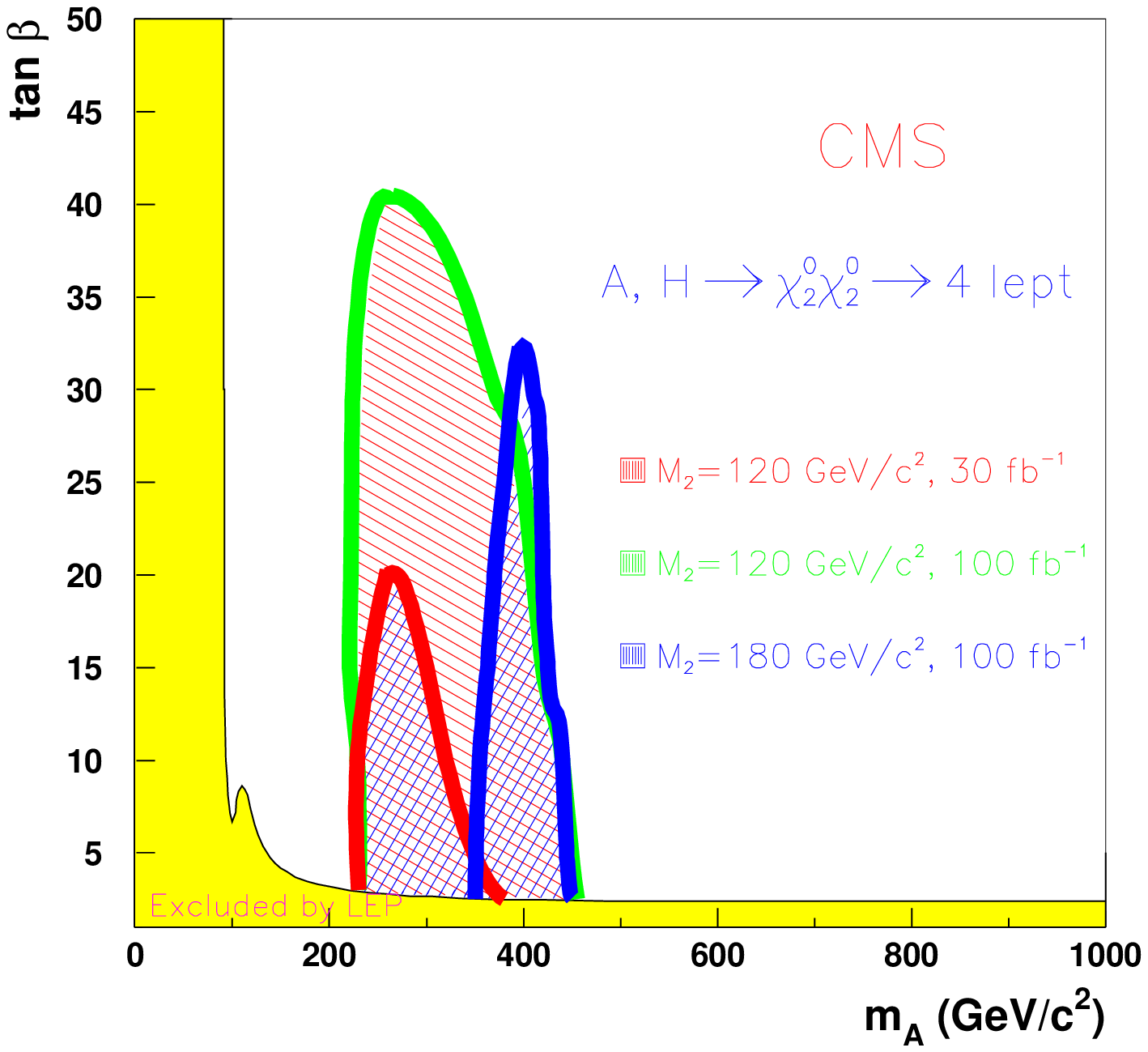,width=0.365\textwidth}
  \put(-310,127){\scalebox{1.1}[1.1]{a)}}
  \put(-132,127){\scalebox{1.1}[1.1]{b)}}
  \vspace*{-0.5mm}
  \caption{a) Upper limit on the branching ratio for ${\rm h \rightarrow}$ invisible with ${\rm 10\ fb^{-1}}$. b) 5 ${\rm \sigma}$-discovery contours for ${\rm A/H \rightarrow \chi^0_2 \chi^0_2 \rightarrow \ell^\pm + X}$. Two integrated luminosities and two sets of supersymmetric parameters are assumed.
  \label{fig:invisible}}
\end{figure}

Higgs bosons decaying to undetectable and therefore invisible particles can be found never- theless~\cite{ref:cms_vbf}, if the Higgs boson is produced in association with other particles that can be seen in the detector. Production through vector boson fusion yields the Higgs boson recoiling against the two jets. In the case of an invisible Higgs boson decay, this final state leads to missing transverse energy and two distinct jets in the detector, such that the event triggers the data acquisition. The exclusion curve of the branching ratio to invisible particles at 95\% confidence level is shown in Fig.~\ref{fig:invisible}a. A natural candidate for invisible decays is the lightest supersymmetric particle, stable in the case of R-parity conservation, such as the lightest neutralino ${\rm \chi^0_1}$.

Another interesting channel is ${\rm A/H \rightarrow \chi^0_2 \chi^0_2 \rightarrow \tilde{\ell^+}\ell^- \tilde{\ell^-}\ell^+ \rightarrow \ell^+ \ell^- \chi^0_1 \ell^+ \ell^- \chi^0_1}$ with four leptons and missing transverse energy from two neutralinos in the final state~\cite{ref:cms_vbf}. This kind of signal, also typical of supersymmetry in general, is almost background free and serves as another interesting possibility for discovering a Higgs boson in an exotic scenario, as shown in Fig.~\ref{fig:invisible}b.

\section{Conclusions}
A selection of mainly new studies has been presented. Although this overview is far from being complete, it gives an idea of the whole picture of Higgs physics at the LHC.

\begin{itemize}
\item The Standard Model Higgs boson will be probed in the entire mass range allowed by vacuum stability and triviality arguments. Good theoretical and experimental understanding helps to improve the corresponding searches. Among a large variety of possible final states, the channels in which Higgs bosons are produced in weak boson fusion allow the backgrounds to be controlled through jet tagging.
\item The mass of the Higgs boson can be measured with high precision. In addition, it is possible to measure also other parameters with a lower precision. These measurements allow the Higgs boson nature to be tested and provides the possibility to distinguish between different models.
\item There is a huge number of possible models. Although it is impossible to study each single case, a set of benchmark scenarios has been picked out. As an example for supersymmetric models, the MSSM with its five Higgs bosons has been studied in detail. Other benchmark scenarios are Higgs bosons produced in decays of heavier particles or Higgs bosons decaying to invisible particles.
\end{itemize}

\noindent
All in all, Higgs physics at the LHC is a rich field. The prospects for various searches and measurements are very promising, but the time to draw final conclusions has not yet come.
\\\

\section*{Acknowledgments}
Work supported in part by the European Community's Human Potential Programme under contract HPRN-CT-2002-00326, [V.D.].
Many thanks to all contributors from ATLAS and CMS and to the organizers of this conference!
\\\

\end{document}